\documentclass[pdf]{UoBnote}
\usepackage[usenames,dvipsnames]{color}     
\usepackage{verbatim}
\usepackage{url}
\usepackage[latin1]{inputenc}
\usepackage[T1]{fontenc}
\usepackage{ae}
\usepackage{amsmath, amssymb}
\usepackage{subfigure}
\usepackage{xspace}
\usepackage{multicol}

\usepackage{finesse}

\usepackage{fancyvrb}
\DefineVerbatimEnvironment{finesse}{Verbatim}
{formatcom=\small}

\author{Charlotte Bond, Daniel Brown and Andreas Freise}
\shorttitle{Gravitational wave signals in \Finesse}
\title{Interferometer responses to gravitational waves:
\\
Comparing \Finesse simulations and analytical solutions}
\date{\today}
\issue{1}
\ligodcc{T1300190}

\usepackage[pdftex,a4paper,pagebackref=true,pdfpagelabels=true]{hyperref}
\definecolor{linkcolor}{rgb}{.8,0,0}
\definecolor{urlcolor}{rgb}{0,0,.7}
\definecolor{citecolor}{rgb}{0,.5,0}
\definecolor{acrocolor}{rgb}{0,0,.7}
\hypersetup{bookmarksopen,colorlinks=true}
\hypersetup{pdfstartview=FitH}
\hypersetup{linktocpage=true,bookmarksnumbered=true}
\hypersetup{plainpages=false,breaklinks=true}
\hypersetup{linkcolor=linkcolor,citecolor=citecolor,urlcolor=urlcolor}

\begin{document}
\maketitle
\tableofcontents
\vspace{1cm}\hrule \vspace{1cm}

\section{Introduction}
This note shows a comparison of analytic calculations and \Finesse
\cite{freise2004frequency} simulations of interferometer responses to
gravitational wave strain. \Finesse includes the possibility to model
gravitational wave signals by modulating the `space' between optical
components. For the validation of the code we could not find
an easily available document showing example responses for various
interferometer types. Thus in this document we present
the analytical results for several simple interferometers
and show that \Finesse gives the same results. This document
should provide useful examples for other people who
find themselves looking for a reference calculation.

\section{Phase modulation in the sideband picture}
Generally we can describe a light field at a given point:
\begin{equation}
E_{\mathrm{in}} = E_0 \exp{(iw_0t + \varphi_0)}
\end{equation}
where $\varphi$ is a constant phase term.  Applying a phase modulation we get:
\begin{equation}
E_{\mathrm{out}} = E_0 \exp{(i (w_0t + \varphi_0 +  \phi(t)))}
\end{equation}
where:
\begin{equation}
\phi(t) = m\cos{(\Omega t + \varphi_s)}
\end{equation}
$m$ is the modulation index and $\varphi_s$ is the modulation signal's phase.  $E_{out}$ can then
be expanded as a series of Bessel functions of the first kind, $J_k(m)$:
\begin{equation}
\label{eq:bessel0}
\exp(\I m \cos\varphi)=\sum_{k=-\infty}^\infty \I^kJ_k(m)\exp(\I k \varphi),
\end{equation}
This implies the creation of an infinite number of upper ($k>0$) and lower ($k<0$)
sidebands around the carrier ($k=0$).  For small modulation indices ($m<1$) 
the Bessel functions decrease 
rapidly with increasing $k$ and so we can use the approximation:
\begin{equation}
J_k(m)~=\left(\frac{m}{2}\right)^k\sum_{n=0}^\infty\frac{\left(-\frac{m^2}{4}\right)^n}
{n! (k+n)!}=\frac{1}{k!}\left(\frac{m}{2}\right)^k+O\left(m^{k+2}\right).
\end{equation}

For $m \ll 1$, as is the case for modulation by a gravitational wave, we can 
express the phase modulation as the addition of two sidebands at frequencies 
$w_0\pm\Omega$ ($k=\pm1$) and a small correction to the amplitude of the carrier ($k=2$):
\begin{equation}
   \begin{split}
      E_{\mathrm{out}} = {}& E_{0} \left(1-\frac{m^2}{4}\right) \exp{( i(w_0t + \varphi_0)} \\
         {}& + E_{0} \frac{m}{2} \exp{\left(i\left((w_0-\Omega)t + \varphi_0 + \frac{\pi}{2}- \varphi_s\right)\right)}  \\
         {}& + E_{0} \frac{m}{2} \exp{\left(i\left((w_0+\Omega)t + \varphi_0 + \frac{\pi}{2} + \varphi_s\right)\right)}  \\
   \end{split}
\end{equation}
where the first term is the carrier, the second term is the lower sideband and the third term the upper sideband.  Hence we have sideband amplitudes of:
\begin{equation}\label{eq:mod_amp}
A_{sb} = \frac{m}{2}\,E_0
\end{equation}
and sideband phases of:
\begin{equation}\label{eq:mod_phs}
\varphi_{sb} = \varphi_0 + \frac{\pi}{2} \pm \varphi_s
\end{equation}
where $\varphi_0$ is the phase of the carrier and $\varphi_s$ is the phase of the modulation signal.

\section{Modulation of a space by a gravitational wave}

\begin{figure}[htb]
\centering
	\includegraphics[scale=0.51, viewport= 80 400 540 730]{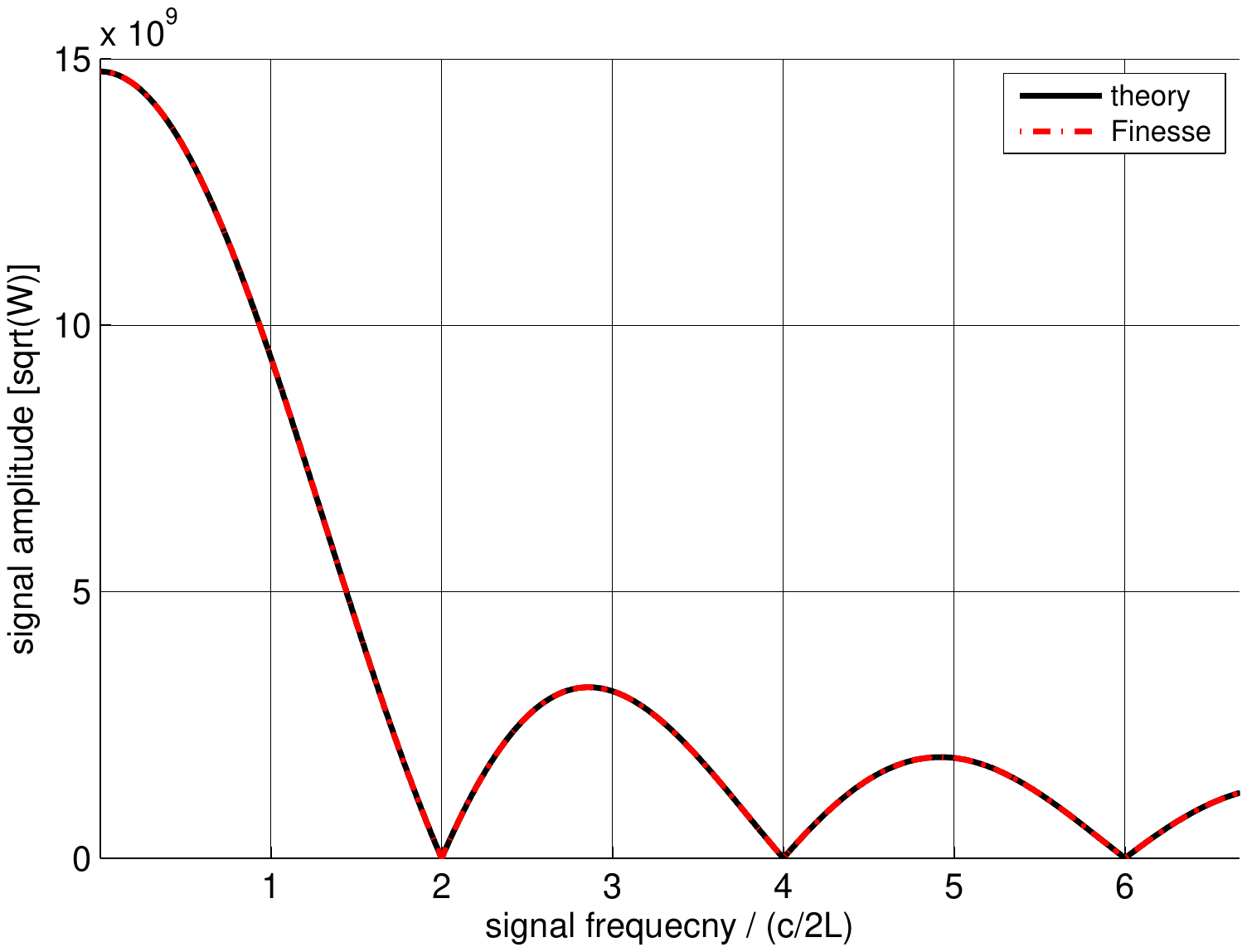}
	\includegraphics[scale=0.51,viewport= 80 400 540 730]{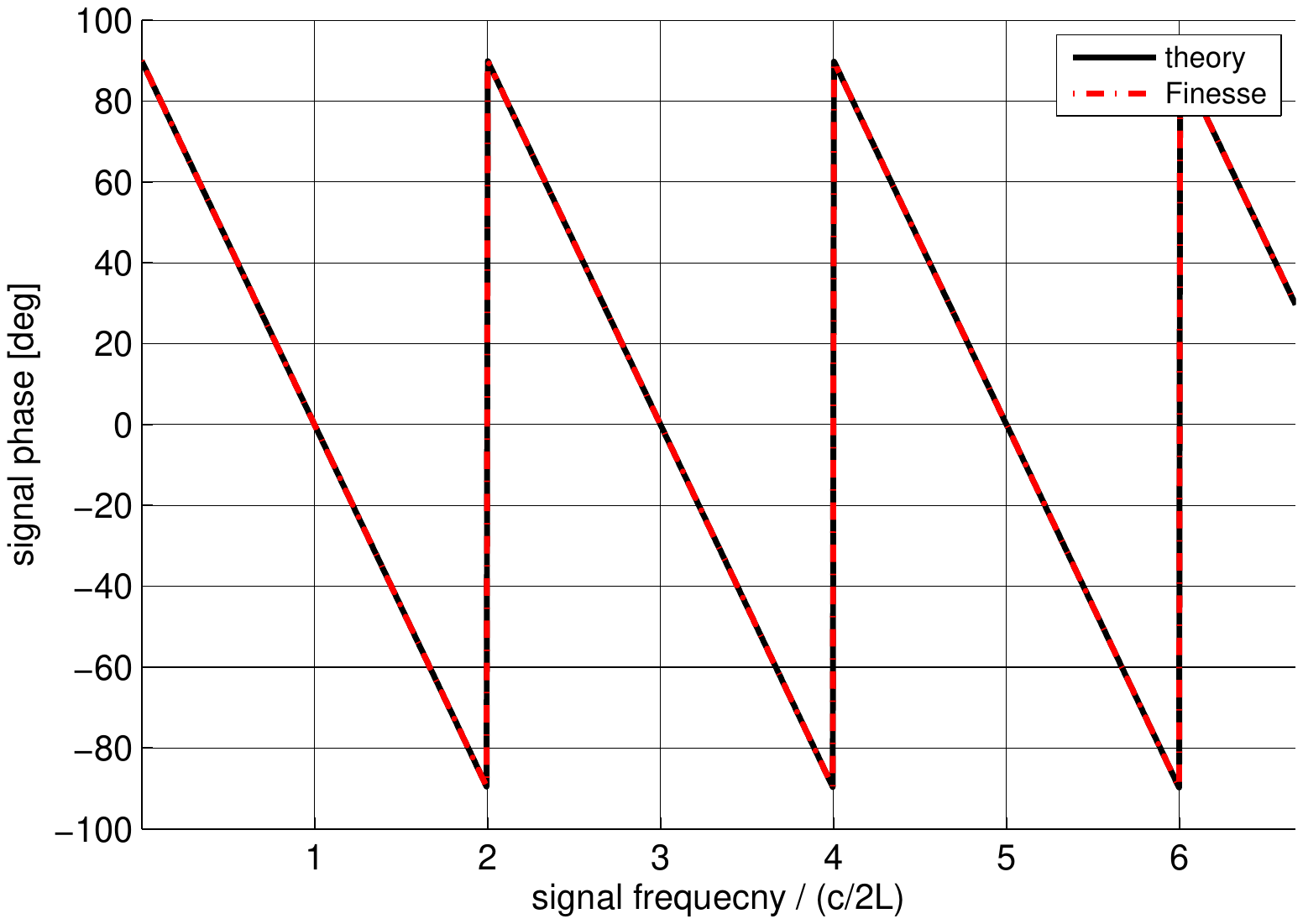}
\caption{Plots showing the amplitude and phase of the upper sideband produced when a gravitational wave modulates a space of length $L=10$\,km against the signal frequency of the gravitational wave.  The signal frequency is normalised with respect to the of the light round-trip of space $L$, or the \emph{free-spectral-range} of a cavity of length $L$.}
\label{fig:space_mod}
\end{figure}

A gravitational wave modulates the length of a space.  In~\cite{Mizuno} the phase change for a round trip between two test masses separated by length $L$ is given by:
\begin{equation}
\varphi(t) = \frac{2 w_0 L}{c} \pm \frac{w_0}{2} \int_{t-2L/c}^{t}h_+(t) \mathrm{d}t
\end{equation}
As stated here the equation refers to a round trip between two points separated by length $L$.  For the phase change for a one-way trip between the two points, and adjusting to our definition of the phase accumulated between two points ($\exp{(-ikL)}$), we have:

\begin{equation}
\varphi=-\frac{\w_0 L}{c} \mp \frac{\w_0}{2}\int_{t-L/c}^{t} h(t) =
-\frac{\w_0 L}{c}\mp \delta \varphi
\end{equation}
We assume we have a gravitational wave signal:
\begin{equation}
h(t)=h_0\cos\left( \w_g t + \varphi_g\right)
\end{equation}
where $\w_g$ and $\varphi_g$ are the user-defined frequency and phase of
the gravitational wave. Thus we get:
\begin{equation}
\begin{array}{lll}
\delta \varphi &=& \frac{\w_0 h_0}{2}
\left[\frac{1}{\w_g}\sin\left(\w_g t
    +\varphi_g\right)\right]_{t-L/c}^{t}\\
&=& \frac{\w_0 h_0}{2\w_g}\left(\sin\left(\w_g t +\varphi_g\right) -
  \sin\left(\w_g t - \w_g \frac{L}{c}+\varphi_g\right)\right)\\
\end{array}
\end{equation}
Using the trigonometric identity $\sin u - \sin
v=2\cos((u+v)/2)\sin((u-v)/2)$ we can write:
\begin{equation}
\begin{array}{lll}
&=& \frac{\w_0 h_0}{\w_g}\cos\left(\w_g t +\varphi_g -\w_g\frac{L}{2c}\right)\sin\left(\w_g\frac{L}{2c}\right)
\end{array}
\end{equation}
This represents a phase modulation with an amplitude of 
\begin{equation}
m = -\frac{w_0 h_0}{w_g} \sin{\left(\frac{w_g L}{2c}\right)} 
\end{equation}
and a phase of:
\begin{equation}
\varphi = - \frac{\omega_g L}{2 c} + \varphi_g
\end{equation}
From equations~\ref{eq:mod_amp} and ~\ref{eq:mod_phs} we can state the
amplitude and phase of the generated sidebands as:
\begin{equation}
A_{\rm sb} = -\frac{w_0 h_0}{2w_g} \sin{\left(\frac{w_g L}{2c}\right)} E_0
\end{equation}
and:
\begin{equation}
\varphi_{\rm sb} = \varphi_0 + \frac{\pi}{2}  - \frac{\w_0 L}{c} \pm \varphi_g \mp \frac{w_g L}{2c} 
\end{equation}
\noindent
Figure~\ref{fig:space_mod} shows plots of the amplitude and phase of the
upper sideband for a single space ($L=10$\,km), comparing the
equations above with the actual \Finesse result. The \Finesse output has been created with this simple file:
\begin{finesse}
l l1 1 0 n1
s s1 10k 1 n1 n2
fsig sm s1 1 0
ad upper 1 n2

xaxis sm f lin 1 100k 1000
put upper f $x1
yaxis abs:deg
\end{finesse}
and the `theory' curves have been created in Matlab with the following
function:
\begin{finesse}
%--------------------------------------------------------------------------
% function [Abs] = FT_GW_sidebands(lambda,h0,fsig,L,n)
%
% A function for Matlab which calculates the amplitude of the sidebands
% created when a light beam travels along a path modulated by a
% gravitational wave.
%
% lambda:   Wavelength of carrier light [m]
% h0:       Gravitational wave amplitude
% fsig:     Frequency of the gravitational wave [Hz]
% L:        Length of the path [m]
% n:        Index refection of the medium through which the beam travels
%
% Asb:      Amplitude of the sidebands [sqrt(W)]
%
% Part of the Simtools package, http://www.gwoptics.org/simtools
% Charlotte Bond    07.11.2012
%--------------------------------------------------------------------------
%

function [Asb] = FT_GW_sidebands(lambda,h0,fsig,L,n,sb_sign)
    % Carrier light parameters
    c = 299792458;
    f0 = c/lambda;
    w0 = 2*pi*f0;
    % Signal anglar frequency
    wsig = 2*pi*fsig;
    % Sideband amplitude
    Asb = (w0*h0./(2*wsig)) .* sin(wsig*L*n/(2*c));
    % Phase
    phi_sb = pi/2 - w0*L*n/c - sb_sign * wsig*L*n/(2*c);
     % Final sideband
    Asb = Asb.*exp(1i*phi_sb);
end

\end{finesse}

\section{Reflection from a mirror}

\begin{figure}[b]
\begin{center}
\includegraphics[scale=0.5, viewport=0 0 240 190] {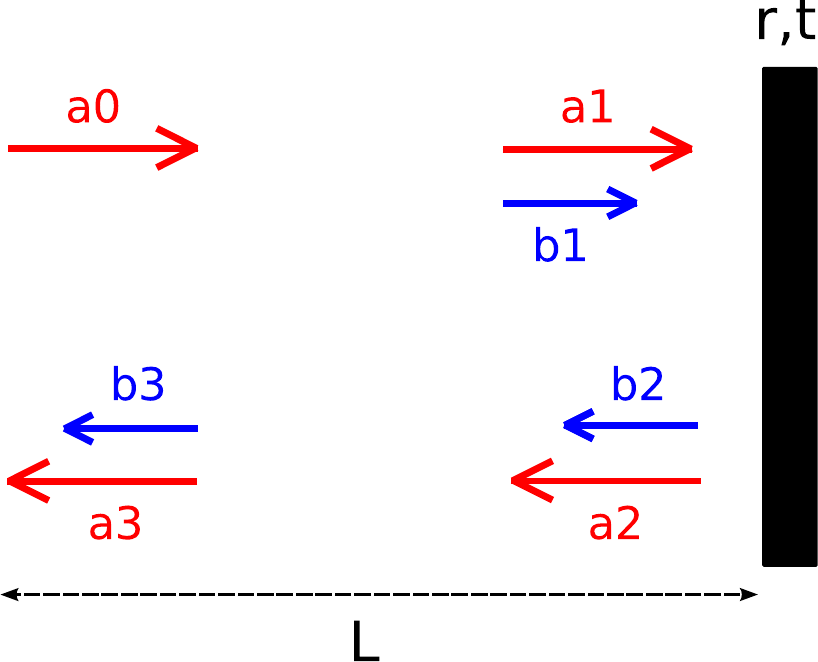}
\caption{A diagram of a single reflection from a mirror.  $a$ represent the carrier field, 
$b$ represent the upper and lower sidebands produced by a gravitational wave.}
\label{fig:single_reflection}
\end{center}
\end{figure}

We now consider the effect of a gravitational wave on a beam propagating through a space of
length $L$ where it is then reflected from a mirror and propagates back through the space
(see figure~\ref{fig:single_reflection}).  \emph{Is this just equivalent to a space of
double the length, taking into account the reflectivity of the mirror?}

\begin{figure}[b]
\centering
	\includegraphics[scale=0.51, viewport= 80 400 540 730]{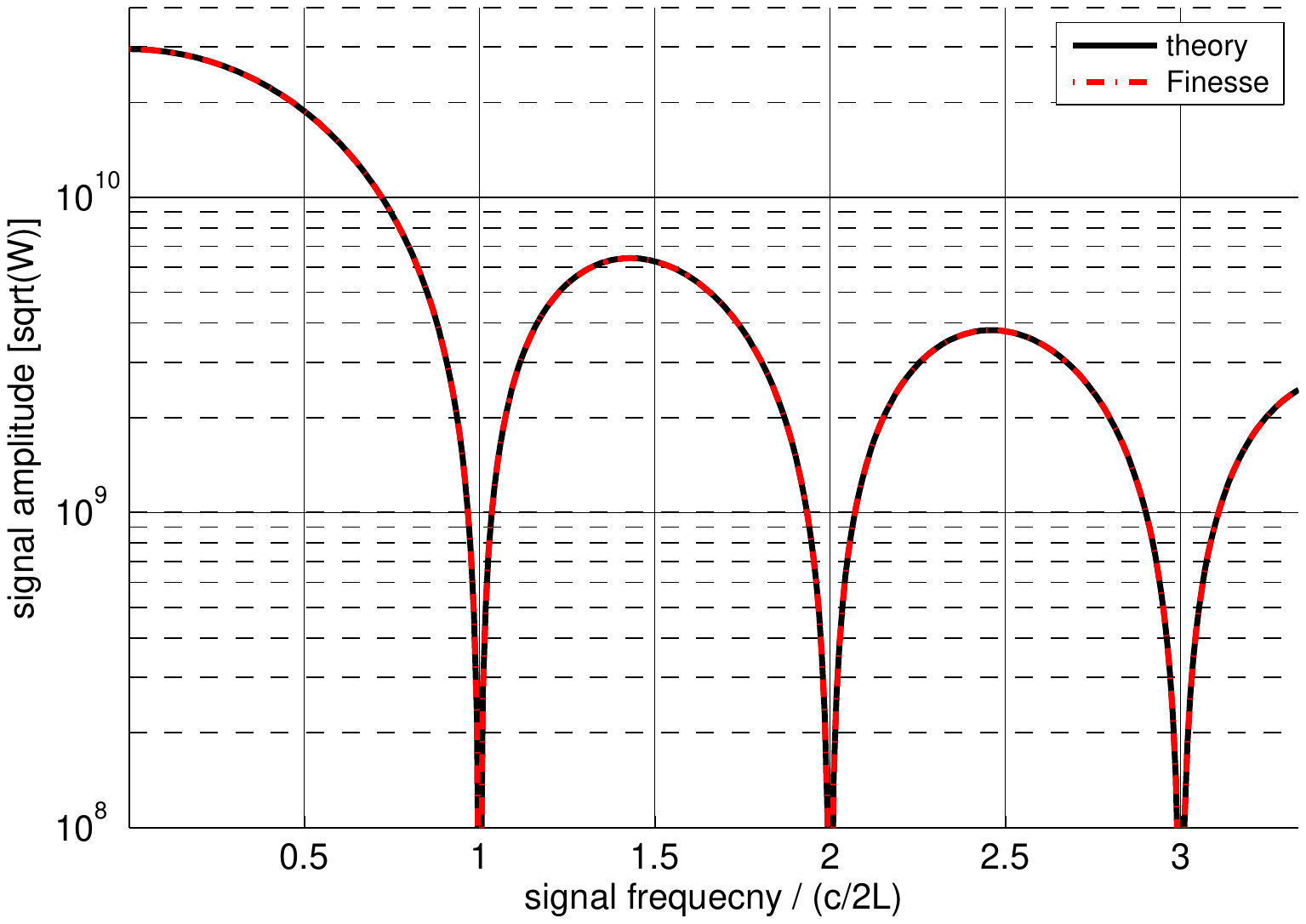}\,
	\includegraphics[scale=0.51,viewport= 80 400 540 730]{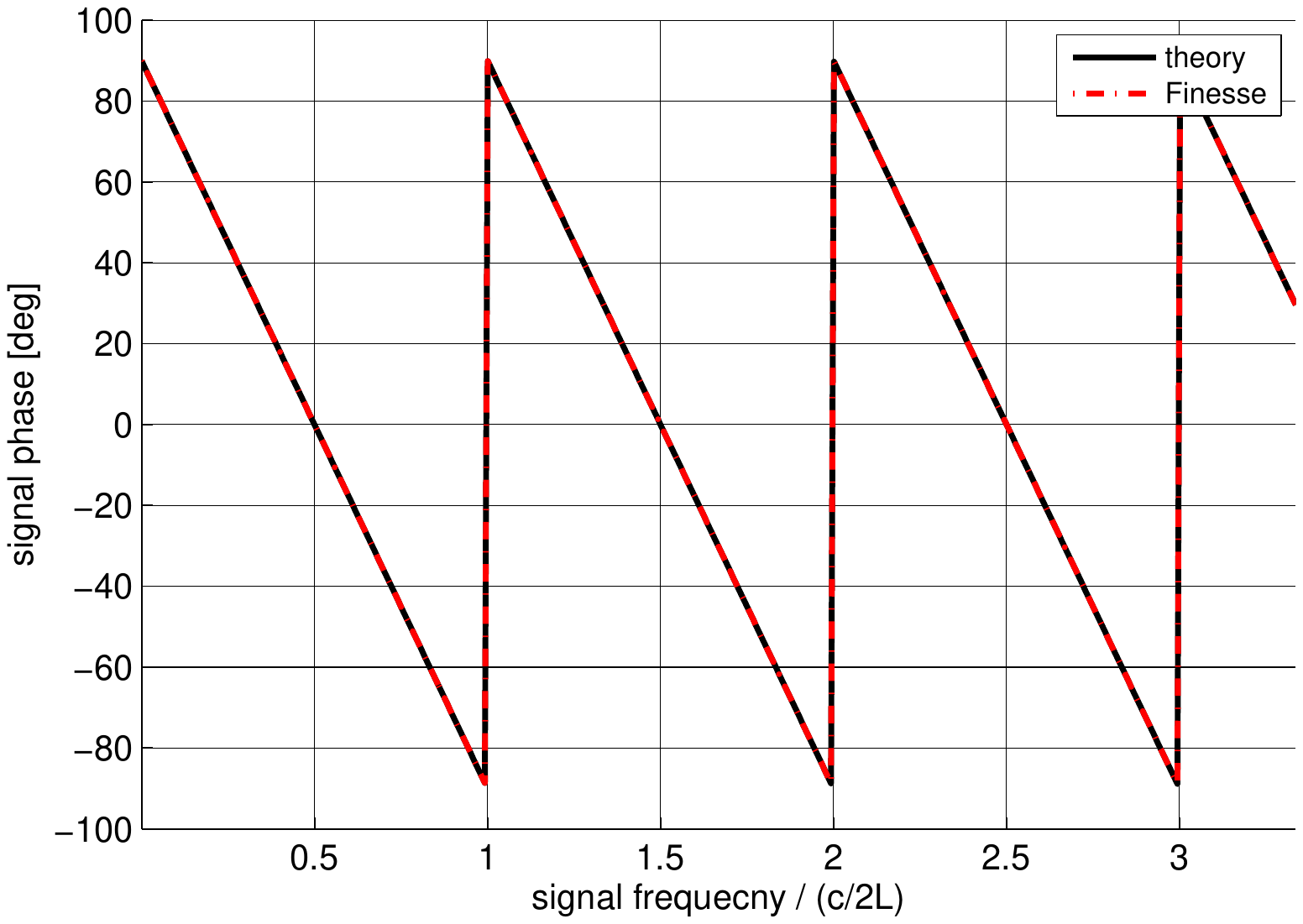}
\caption{Plots showing the amplitude and phase of the upper sideband
  produced when a gravitational wave modulates a carrier field which
  propagates along a space of length $L=10$\,km and is then reflected
  by a mirror ($r=1$) and travels the same $10$\,km back again.  
  This is the same result as shown in 
  Figure~\ref{fig:space_mod} except for the fact
that the $x$-axis is scaled by a factor of two (and in this plot the
$y$-axis uses a log-scale).}
\label{fig:space_mod_reflect}
\end{figure}

\noindent
In this case the effect of the gravitational wave in calculated by considering the sidebands
added at different points in the setup, after each length propagation.  As the modulation
index, $m$, is small we assume the carrier field amplitude is unchanged due to the
gravitational wave.  Referring to the fields in figure~\ref{fig:single_reflection},
where $a$ refers to the field of the carrier and $b$ refer to the field of the sidebands
we have:
\begin{equation*}
   \begin{split}
      {}& a_3 = a_2 \exp{(-ik_0L)}   \\
         {}& a_2 = r a_1 \\
         {}& a_1 = a_0 \exp{(-ik_0L)} 
   \end{split}
\end{equation*}
So the reflected carrier field is given by:
\begin{equation}
a_3 = r a_0 \exp{(-i2k_0L)}
\end{equation}
For the sideband fields we have:
\begin{equation*}
\begin{split}
	{}& b_3 = b_2 \exp{(-i(k_0\pm k_g)L)} + a_2 \ \alpha_{sb}^{\mathrm{space}}\\
	{}& b_2 = r b_1 \\
	{}& b_1 = a_0 \ \alpha_{sb}^{\mathrm{space}} \\
\end{split}
\end{equation*}
where $\alpha_{sb}^{\mathrm{space}}$ describes the relative amplitude and phase of the
sideband created from the modulation of the space.  This gives the reflected field of the
sidebands as:
\begin{equation}
\begin{split}
	b_3 {}& = r a_0 \ \alpha_{sb}^{\mathrm{space}} \exp{(-i(k_0 \pm k_g)L )} + r a_0 \ \alpha_{sb}^{\mathrm{space}} \exp{(-ik_0L )}\\
	{}& =  r a_0 \ \alpha_{sb}^{\mathrm{space}} \exp{(-i k_0 L )} \left[ 1 + \exp{(\mp ik_gL)} \right] \\
\end{split}
\end{equation}
The sidebands produced from the round-trip propagation and single 
reflection have combined amplitude and phase 
$a_0 \ \alpha_{sb}^{\mathrm{arm}}$ where:
\begin{equation}
\alpha_{sb}^{\mathrm{arm}} = r \alpha_{sb}^{\mathrm{space}} \exp{(-ik_0L)} [1+\exp{(\mp ik_gL)}]
\end{equation}
and if we assume the space is `resonant' for the carrier wave we
can simplify this to:
\begin{equation}
\alpha_{sb}^{\mathrm{arm}} = r \alpha_{sb}^{\mathrm{space}} [1+\exp{(\mp ik_gL)}]
\end{equation}
Figure~\ref{fig:space_mod_reflect} shows plots of the amplitude and phase of the upper
sideband for propagation back-and-forth from a mirror ($L=10$\,km, $r=1$), comparing these
analytical equations and the result from \Finesse.  
The \Finesse output is generated by the following commands:
\begin{finesse}
l l1 1 0 n1
s s1 10k 1 n1 n2
m m1 1 0 0 n2 n3
fsig sm s1 1 0
ad upper 1 n1

xaxis sm f lin 1 50k 400
put upper f $x1
yaxis abs:deg
\end{finesse}
The plots illustrate that this propagation back-and-forth is equivalent to the 
modulation of a space of double the length (the plots are identical to those 
shown in figure~\ref{fig:space_mod} except the $x$-axis is scaled by 2).

\section{Linear cavities}
\begin{figure}[htdp]
\begin{center}
\includegraphics[scale=0.5, viewport=120 0 240 190] {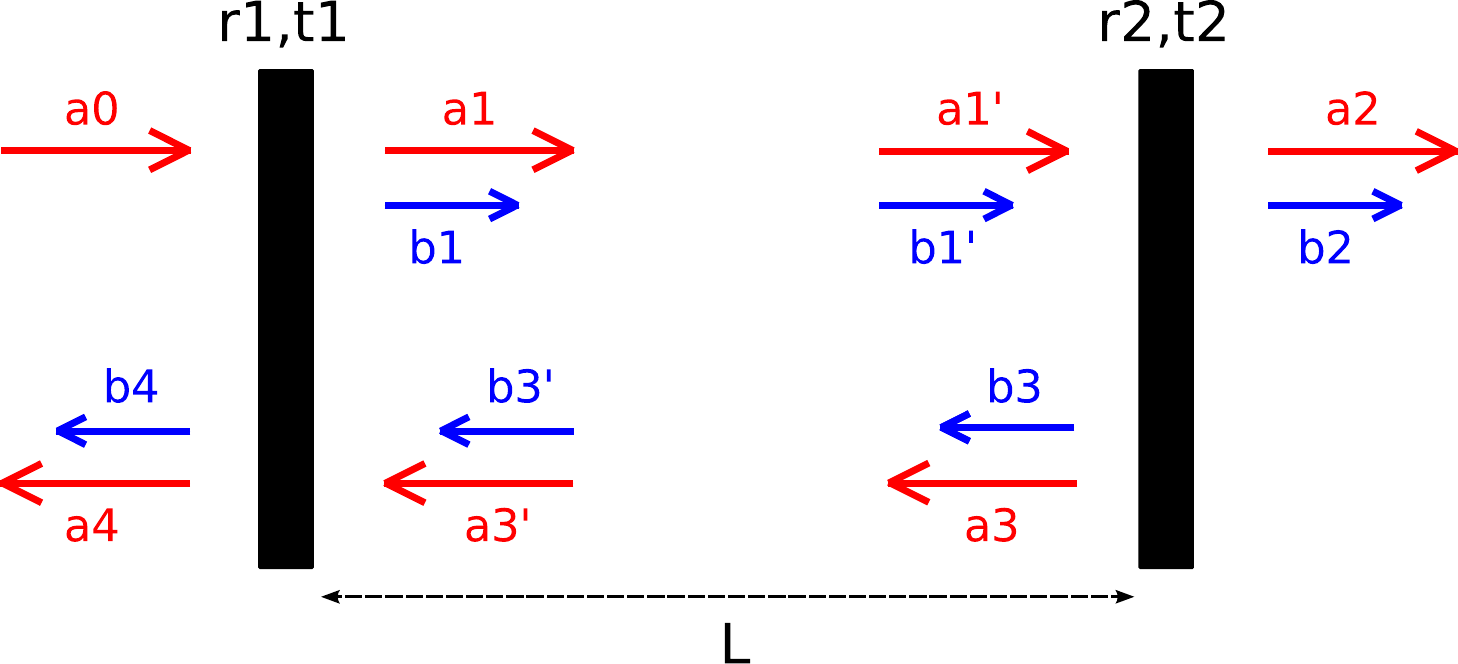}
\caption{A diagram showing the carrier and sideband fields at different points in a linear
cavity.  $a$ represent the carrier field, $b$ represent the upper and lower sidebands 
produced by a gravitational wave.}
\label{fig:cavity}
\end{center}
\end{figure}

We now consider the sidebands reflected from a Fabry-Perot cavity when the cavity space
is modulated by a gravitational wave.  Figure~\ref{fig:cavity} shows the different fields
at different 
points in a linear cavity.

\noindent
The sideband field reflected from a linear cavity is:
\begin{equation}
b_4 = it_1 b_3'
\end{equation}
where
\begin{equation*}
\begin{split}
{}& b_3' = a_1 \alpha_{sb}^{\mathrm{arm}} + r_2b_1\exp{(-i2(k_0\pm k_g)L)} \\
{}& b_1 = r_1b_3' \\
{}& b_3' = \frac{a_1 \alpha_{sb}^{\mathrm{arm}}}{1-r_1r_2\exp{(-i2(k_0\pm k_g)L)}}
\end{split}
\end{equation*}
and $\alpha_{sb}^{\mathrm{arm}}$ refers to the relative amplitude and phase of 
the sidebands after 
propagation back-and-forth from the end mirror.
The carrier fields are solved by the usual simultaneous equations:
\begin{equation*}
\begin{split}
{}& a_1 = it_1 a_0 + r_1 a_3' \\
{}&  a_3' = a_3 \exp{(-ik_0L)} \\
{}& a_3 = r_2 a_1' \\
{}& a_1' = a_1 \exp{(-ik_0L)} 
\end{split} 
\end{equation*}  
from which we have:
\begin{equation}
\begin{split}
{}& a_1 = it_1a_0 + r_1  r_2 a_1 \exp{(-i2k_0L)} \\
{}& a_1 = \frac{it_1a_0}{1-r_1r_2\exp{(-i2k_0L)}}
\end{split}
\end{equation}

\begin{figure}[h]
\centering
	\includegraphics[scale=0.51, viewport= 80 400 540 730]{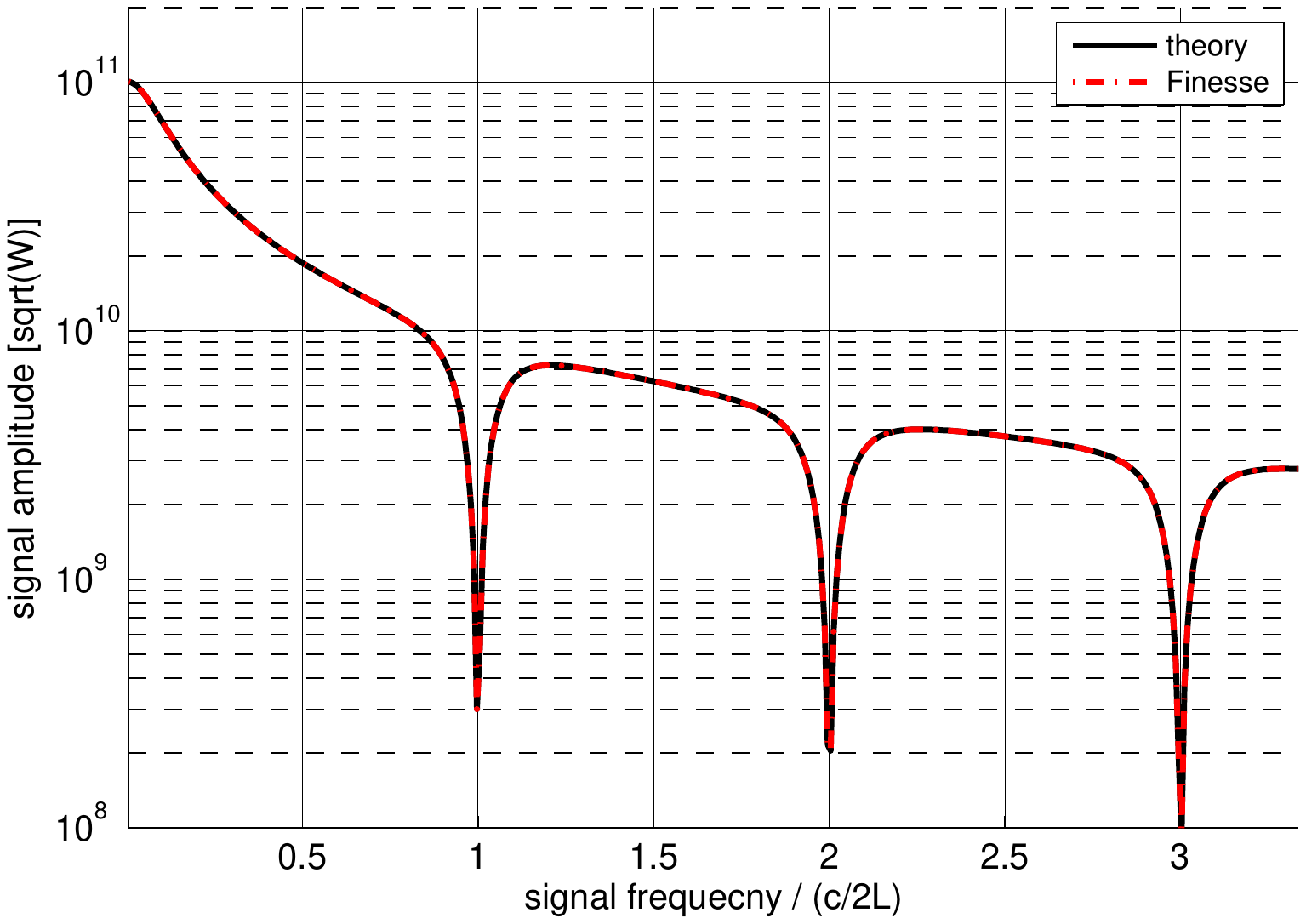}
	\includegraphics[scale=0.51,viewport= 80 400 540 730]{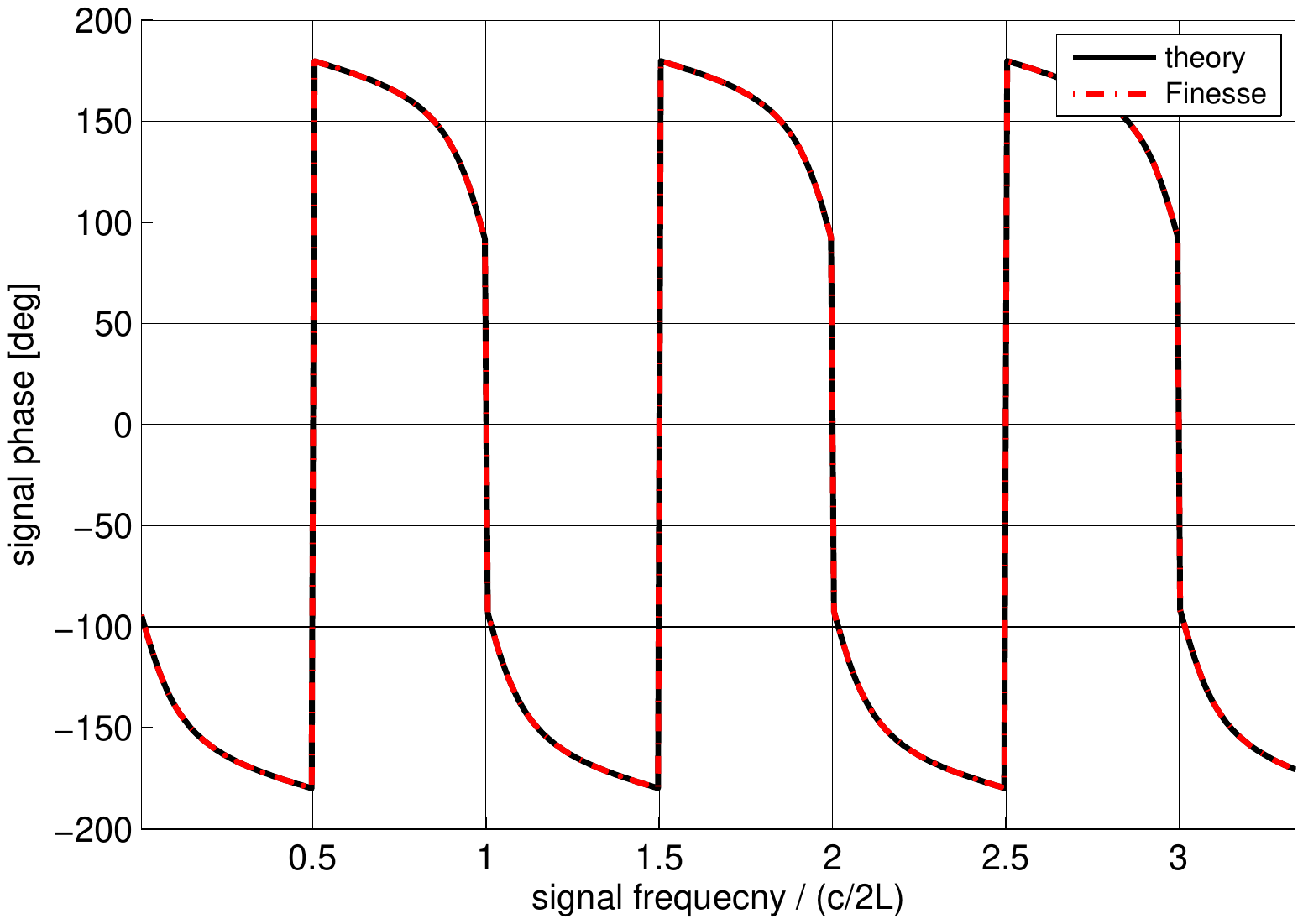}
\caption{Plots showing the amplitude and phase of the upper sideband
  produced by a gravitational wave modulating a Fabry-Perot cavity of
  length $L=10$\,km at frequency against the signal frequency. The sideband is
  detected in the light reflected from the cavity.}
\label{fig:space_mod_cav}
\end{figure}

Finally:
\begin{equation}
\begin{split}
b_4 {}& = \frac{-T_1a_0}{1-r_1r_2\exp{(-i2k_0L)}} \frac{1}{1-r_1r_2\exp{(-i2(k_0\pm k_g)L)}} \alpha_{sb}^{\mathrm{arm}} 
\end{split}
\end{equation} 
The sidebands reflected from a Fabry-Perot cavity are given by the field $a_0 \ \alpha_{sb}^{\mathrm{FP}}$, where:
\begin{equation}
\alpha_{sb}^{\mathrm{FP}} = \frac{-T_1}{1-r_1 r_2 } \frac{\alpha_{sb}^{\mathrm{arm}}}
{1-r_1 r_2 \exp{(\mp i2k_g L)}}
\end{equation}
if we assume the cavity is on resonance.  In figure~\ref{fig:space_mod_cav} plots of this analytic result for a 10\,km long cavity are compared with the result from \Finesse.  The \Finesse output is generated
with the following file:
\begin{finesse}
l l1 1 0 nin

s s0 1 nin n1

const T_ITM 700e-3 
const T_ETM 100e-6 

m1 ITM $T_ITM 0 0 n1 n2
s sarm 10k n2 n3 			
m1 ETM $T_ETM 0 180 n3 n4

fsig sig1 sarm 1 0
ad upper 0 n1

xaxis sig1 f lin 100 50k 400
put upper f $x1
yaxis lin abs:deg
\end{finesse}

\section{Michelson interferometer}

\begin{figure}[t]
\centering
	\includegraphics[scale=0.51, viewport= 80 400 540 730]{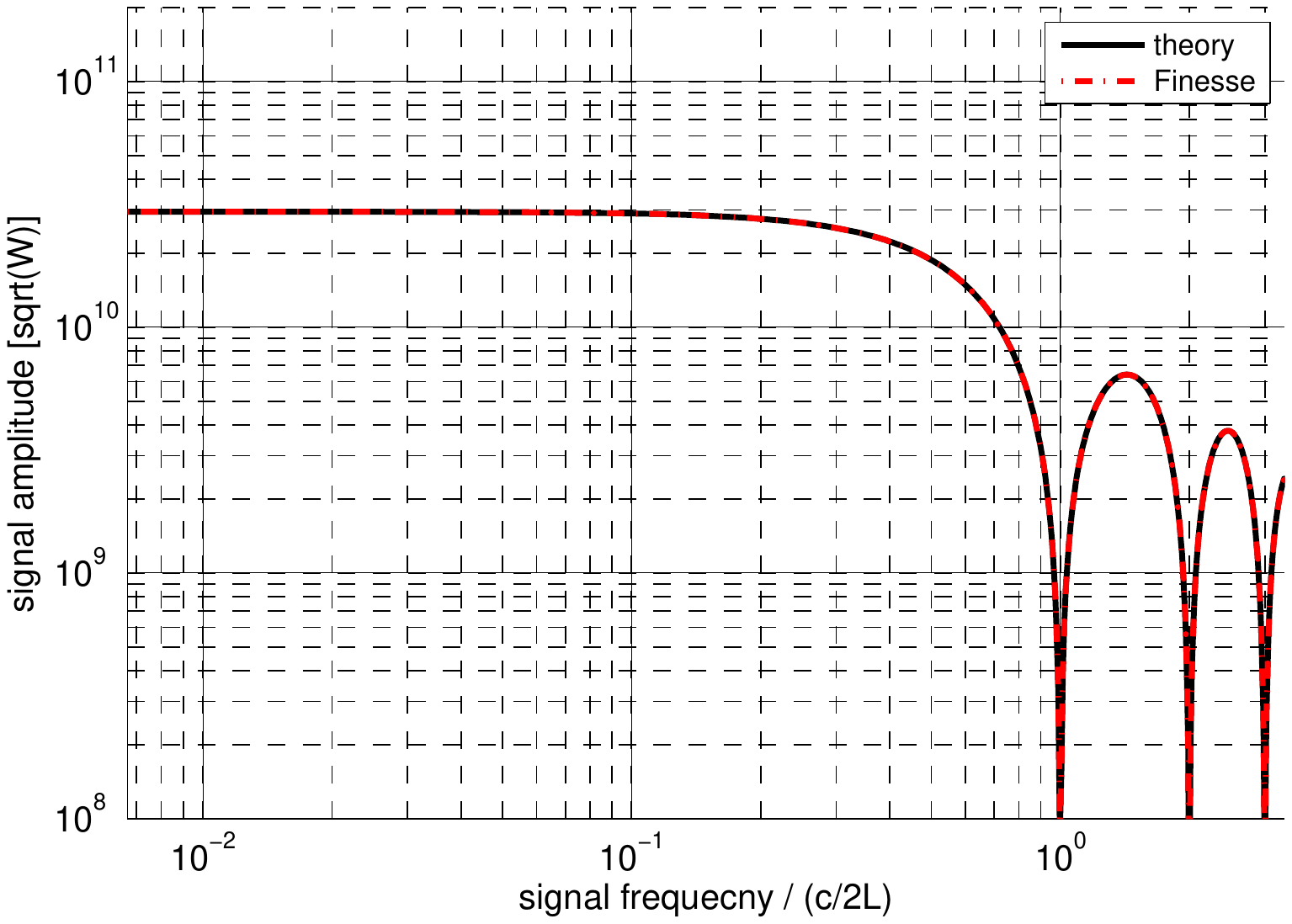}
	\includegraphics[scale=0.51,viewport= 80 400 540 730]{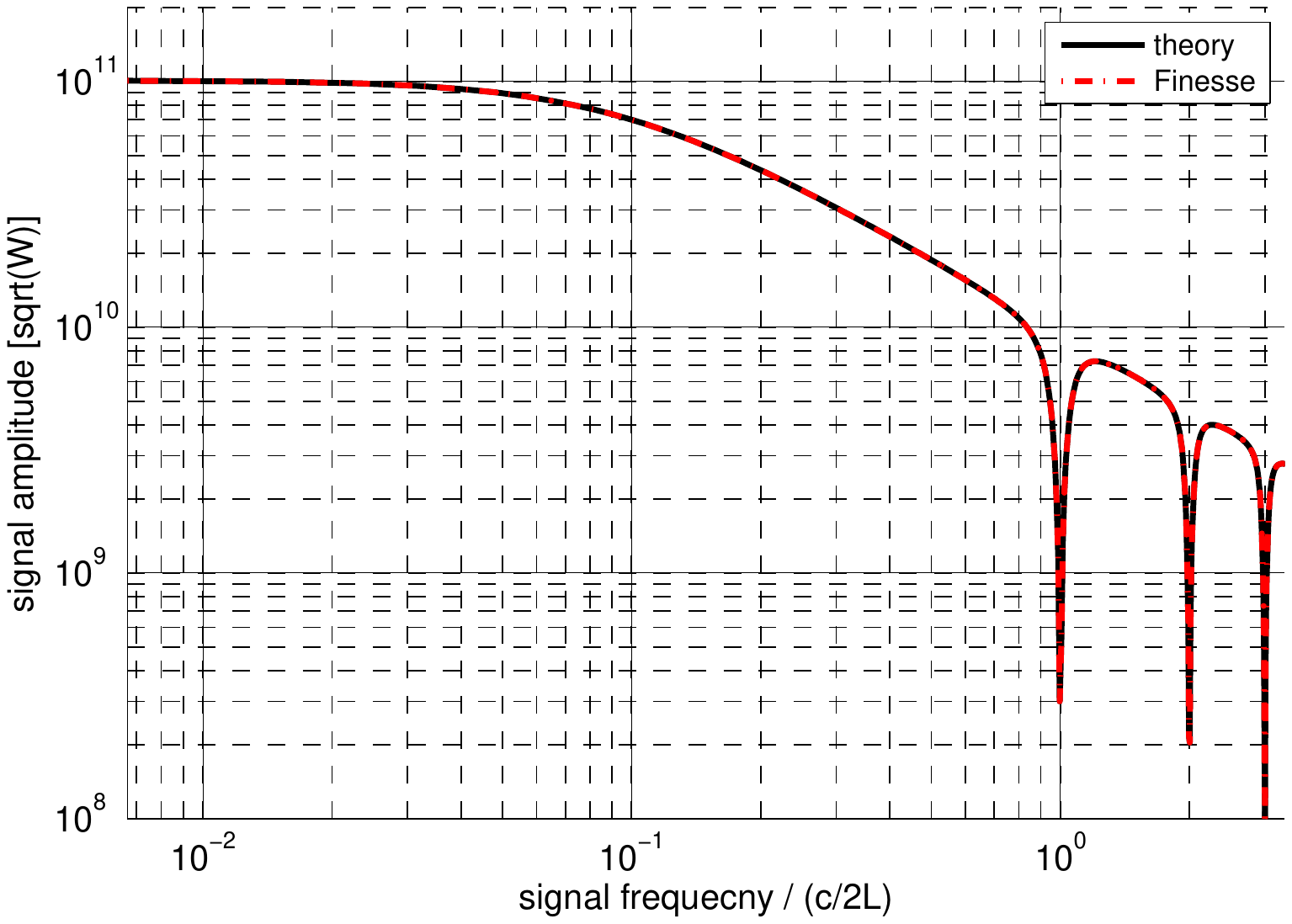} \\
	\includegraphics[scale=0.51, viewport= 80 400 540 730]{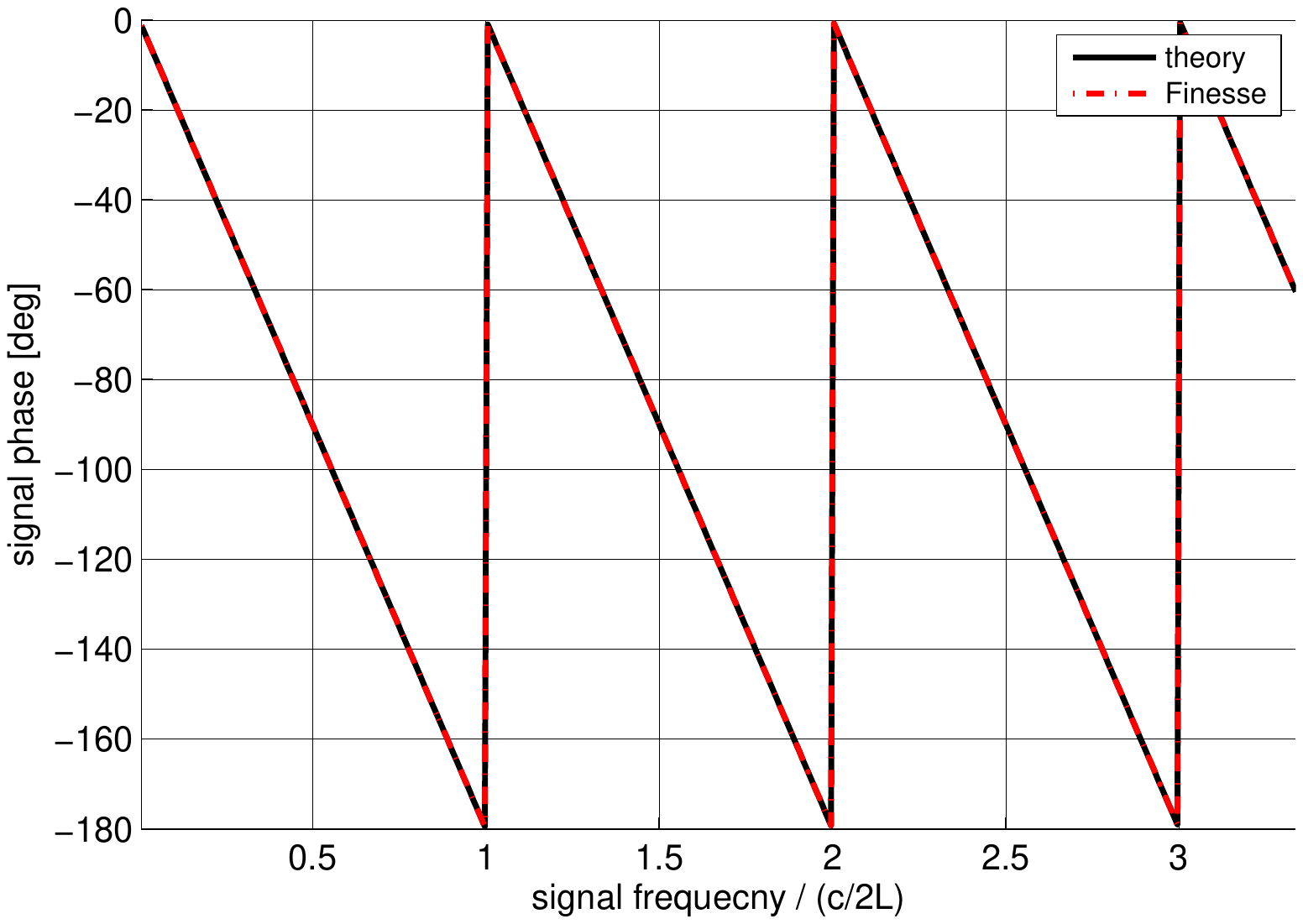}
	\includegraphics[scale=0.51,viewport= 80 400 540 730]{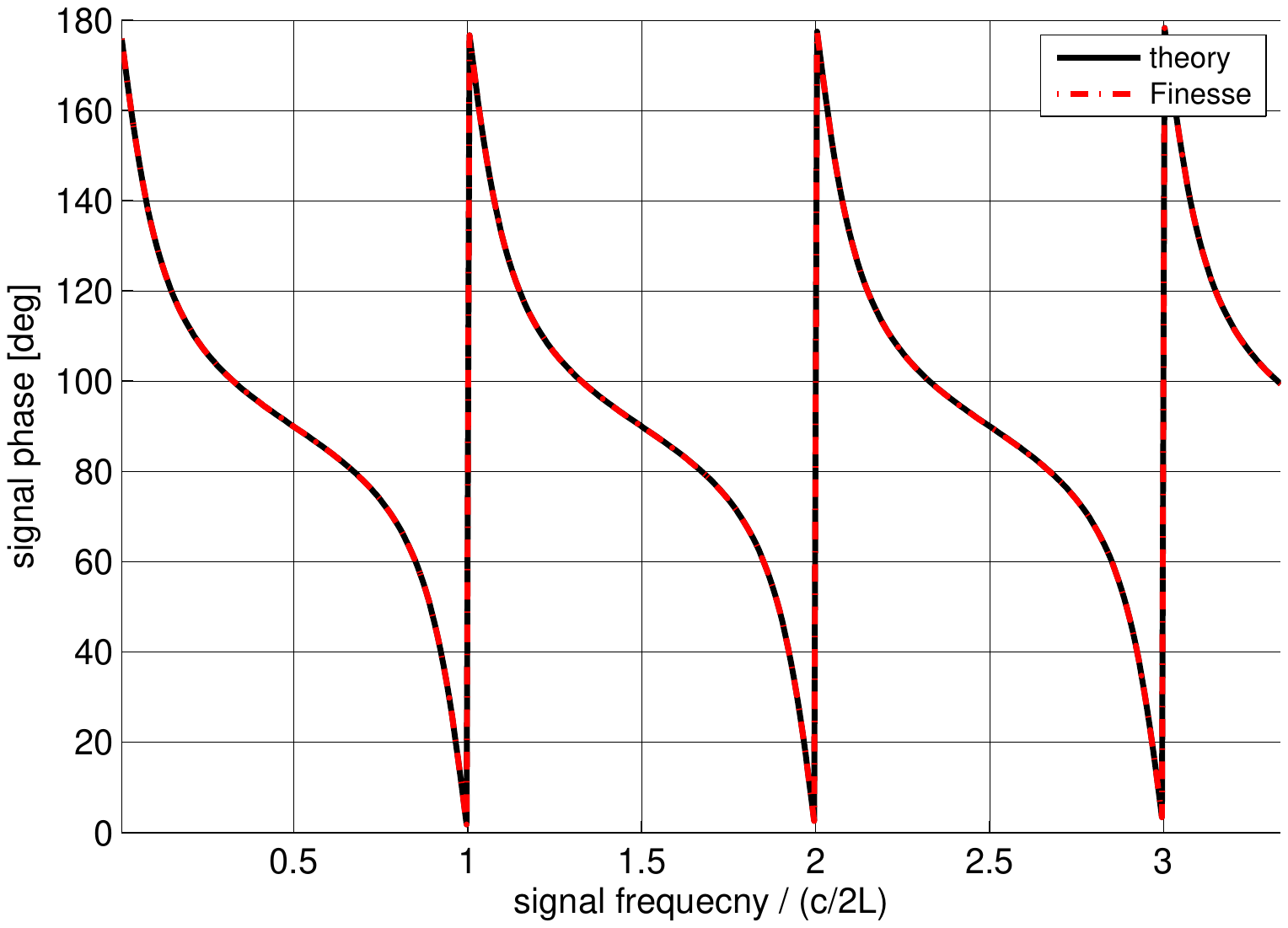}
\caption{Plots showing the amplitude and phase of the upper sideband
  produced by a gravitational wave modulating the 10\,km long arms of a Michelson 
  interferometer.  Left: Plots of the amplitude (top) and phase (bottom) of the 
  sidebands at the output of a simple Michelson with no arm cavities.  
  Right: Plots of the amplitude (top) and phase (bottom) of the sidebands at
  the output of a Michelson with Fabry-Perot arm cavities.}
\label{fig:mod_mich}
\end{figure}

We now look at the effect of a gravitational wave on the output of a Michelson interferometer.  The amplitude of the sidebands at the output of the detector is given by:
\begin{equation}
b_{out} = r_{bs} b_x + i\ t_{bs} b_y
\end{equation}
where $r_{bs}$ and $t_{bs}$ refer to the reflection and transmission coefficients of the beam-splitter
and $b_x$ and $b_y$ are the sideband fields reflected from the $x$ and $y$ arms.  If we consider a gravitational wave in the ideal polarisation for a Michelson (a gravitational wave, $h_+$, modulating
the space in the $y$ arm 180$^{\circ}$ out of phase with the $x$ arm) we have:
\begin{equation*}
\begin{split}
	{}& b_x = i \ t_{bs} \ (a_0 \exp{(-ik_0 l_x)}) \ \alpha_{sb}^{\mathrm{FP}} \exp{(-i(k_0\pm k_g)l_x)} \\
	{}& b_y = r_{bs} \ (a_0 \exp{(-ik_0 l_y)}) \ (-\alpha_{sb}^{\mathrm{FP}}) \exp{(-i(k_0\pm k_g)l_y)}
\end{split}
\end{equation*}
where $l_x$ and $l_y$ refer to the Michelson lengths, which should be much smaller than the cavity lengths.  In order to operate on the dark fringe we must have $|l_x-l_y| = (2N+1)\frac{\lambda}{4}$, where $N$ is an integer.  Finally, at the output of the interferometer we have:
\begin{equation}
	b_{out}  = i \ t_{bs} r_{bs} a_0 \alpha_{sb}^{\mathrm{FP}} \left[ \exp{(-i(2k_0\pm k_g)l_x)} -
	\exp{(-i(2k_0\pm k_g)l_y)} \right]
\end{equation}
For the case of no arm cavities (i.e. just a single mirror at the end of the arm) just replace the $\alpha_{sb}^{\mathrm{FP}}$ factor with $\alpha_{sb}^{\mathrm{arm}}$.  In figure~\ref{fig:mod_mich} this analytic result and the result from a \textsc{Finesse} simulation of the same setup are plotted, for a simple Michelson and a Michelson with arm cavities.  The \textsc{Finesse} output is generated using the following code:

\begin{multicols}{2}{
\noindent
For a simple Michelson without arm cavities:
\begin{verbatim}
l l1 1 0 nin

s s0 1 nin n1

const T_ETM 100e-6 

bs BS 0.5 0.5 0 45 n1 ny1 nx1 nout 

s syarm 10k ny1 ny2 			
m1 ETMy $T_ETM 0 0 ny2 ny3

s sxarm 10k nx1 nx2 			
m1 ETMx $T_ETM 0 90 nx2 nx3

fsig sig1 syarm 1 180
fsig sig1 sxarm 1 0

ad upper 0 nout

xaxis sig1 f lin 100 50k 400
put upper f $x1
yaxis lin abs:deg







\end{verbatim}

\columnbreak

\noindent
For a Michelson with arm cavities:
\begin{verbatim}
l l1 1 0 nin

s s0 1 nin n1

const T_ITM 700e-3 
const T_ETM 100e-6 

bs BS 0.5 0.5 0 45 n1 ny1 nx1 nout 

s sy 1 ny1 ny2

m1 ITMy $T_ITM 0 0 ny2 ny3
s syarm 10k ny3 ny4 			
m1 ETMy $T_ETM 0 0 ny4 ny5

s sx 1 nx1 nx2

m1 ITMx $T_ITM 0 90 nx2 nx3
s sxarm 10k nx3 nx4 			
m1 ETMx $T_ETM 0 90 nx4 nx5

fsig sig1 syarm 1 180
fsig sig1 sxarm 1 0

ad upper 0 nout

xaxis sig1 f lin 100 50k 400
put upper f $x1
yaxis lin abs:deg

\end{verbatim}
}
\end{multicols}

\newpage
 \section{Sagnac}
\begin{figure}[htbp]
\centering
	\includegraphics[scale=0.51, viewport= 80 400 540 730]{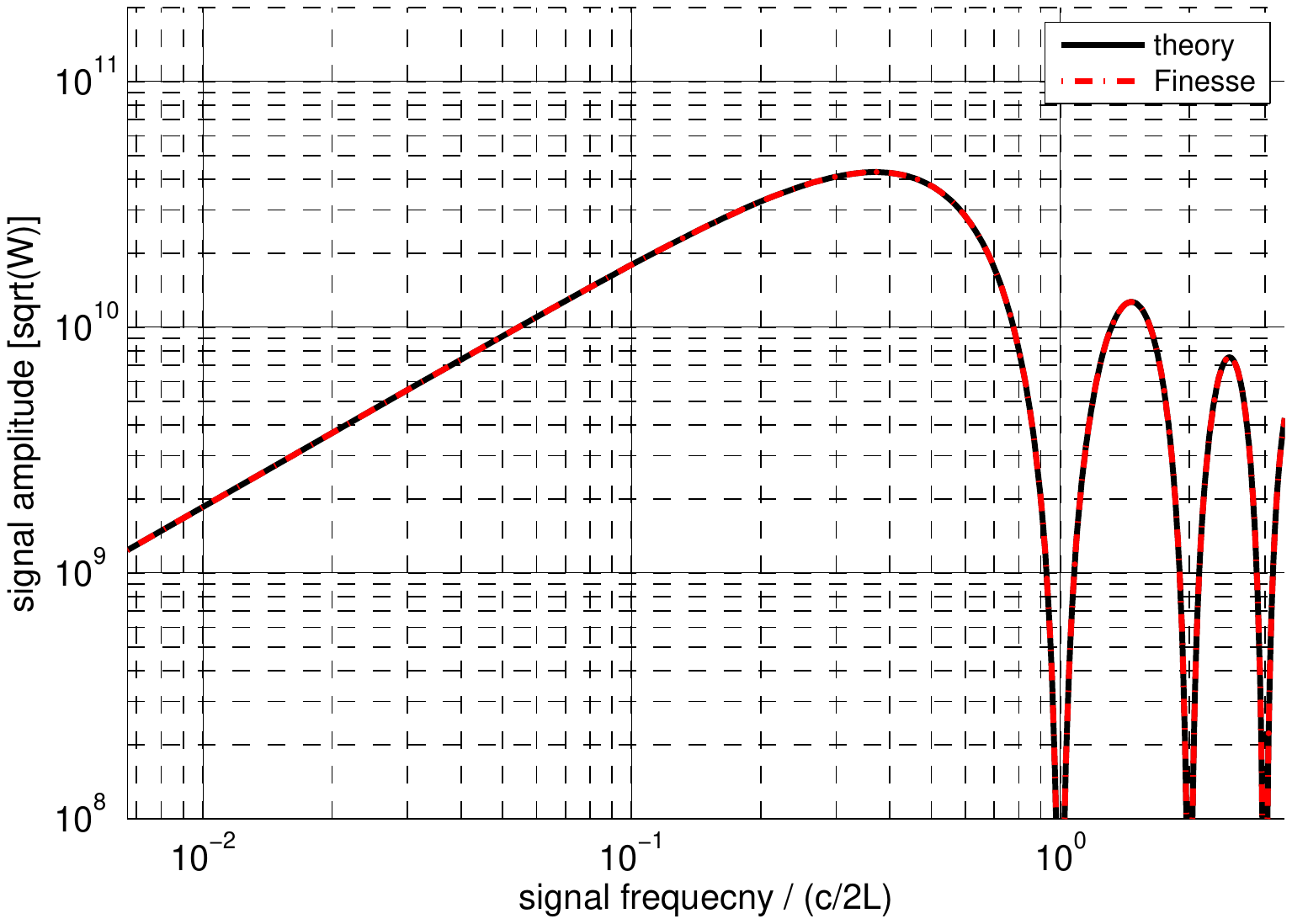}
	\includegraphics[scale=0.51,viewport= 80 400 540 730]{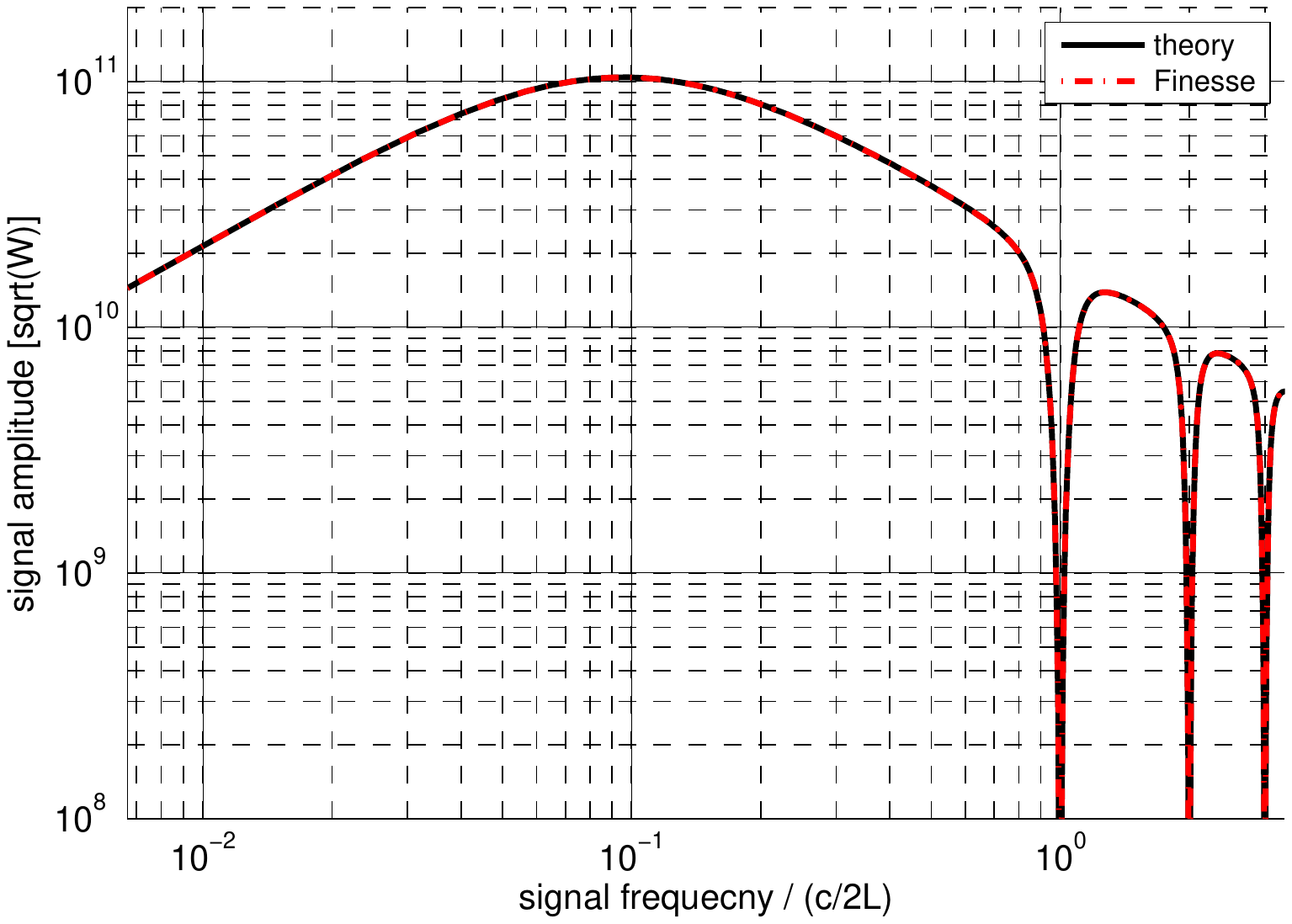} \\
	\includegraphics[scale=0.51, viewport= 80 400 540 730]{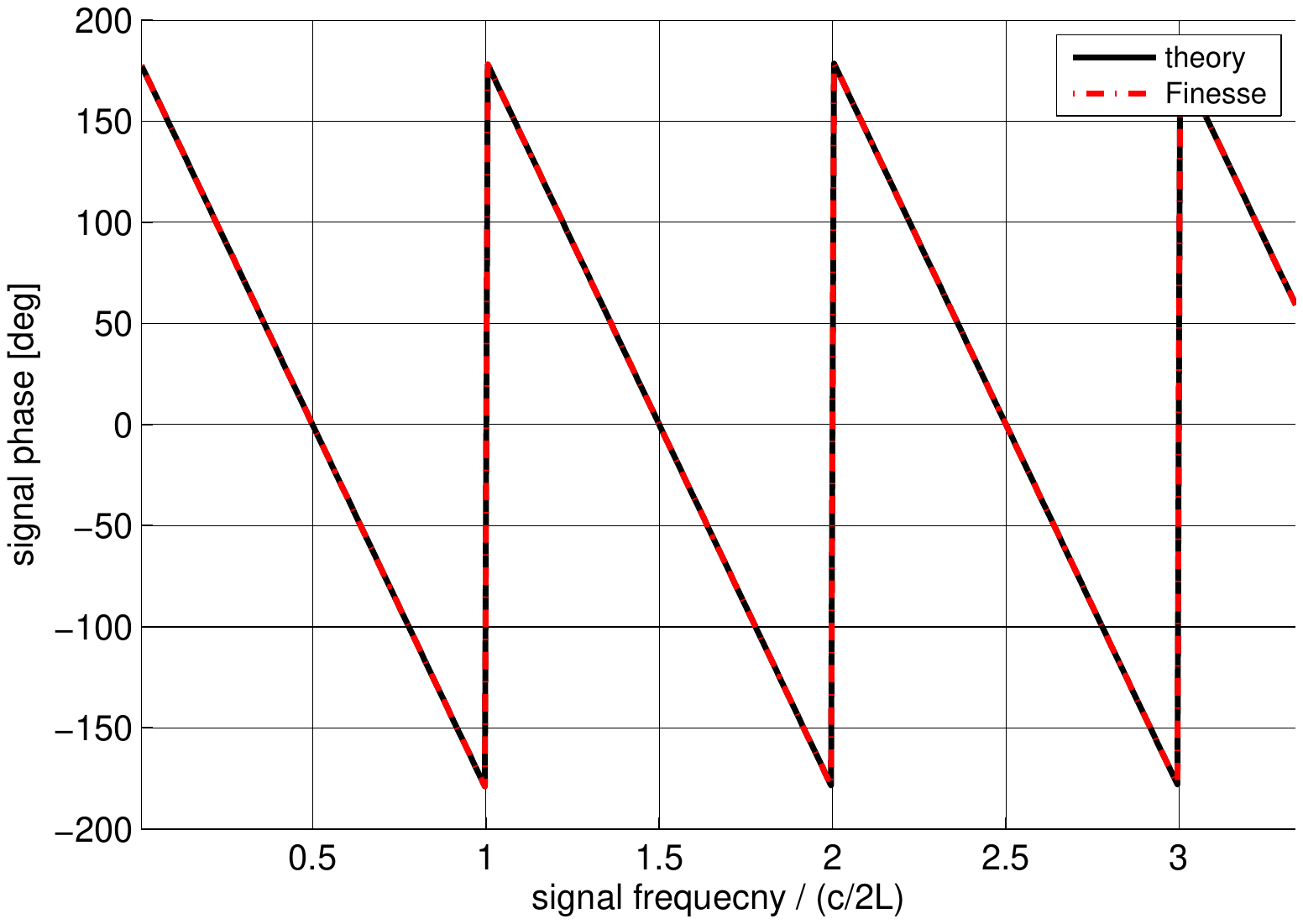}
	\includegraphics[scale=0.51,viewport= 80 400 540 730]{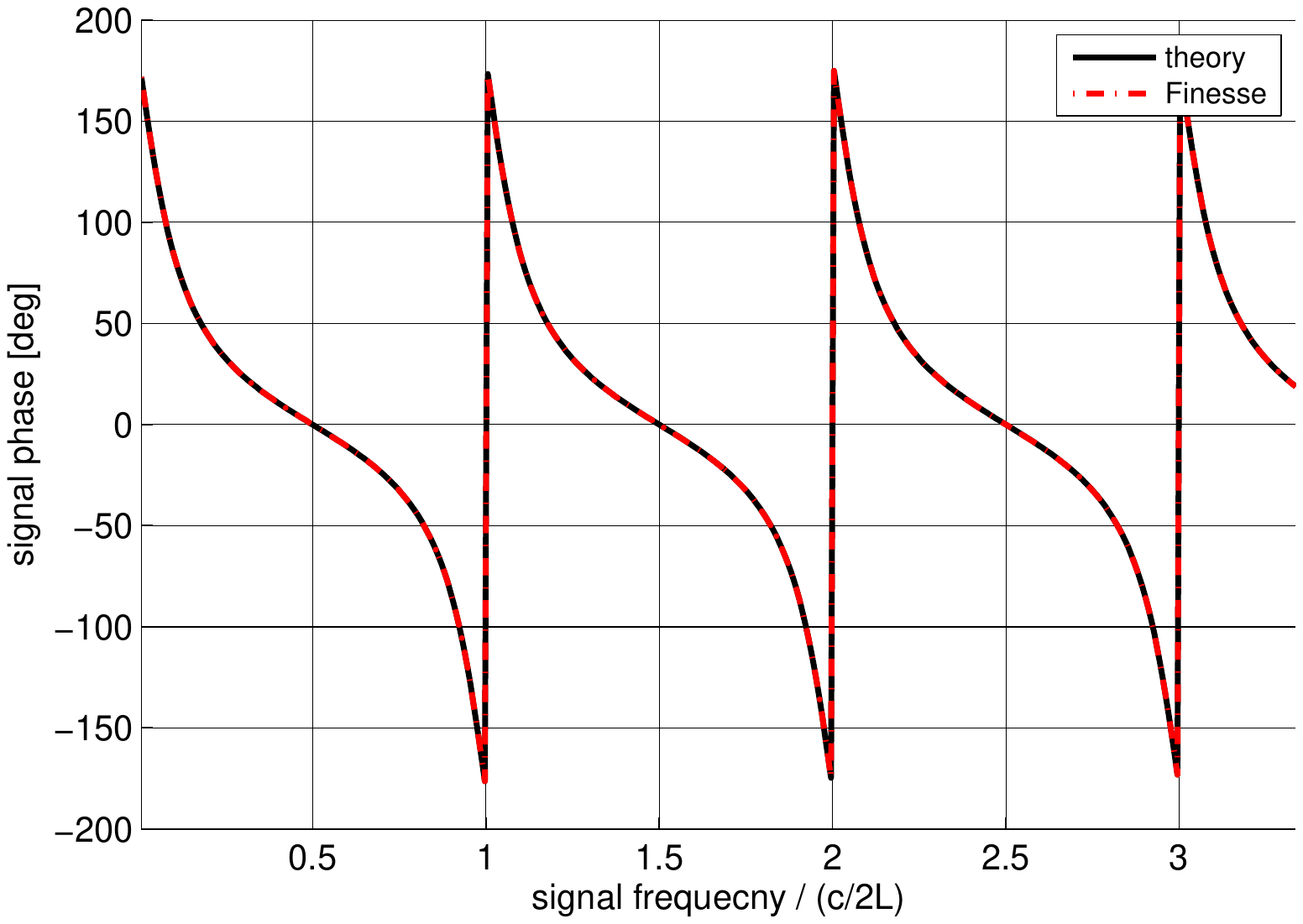}
\caption{Plots showing the amplitude and phase of the upper sideband
  produced by a gravitational wave modulating the arms of a Sagnac 
  interferometer.  Left: Plots of the amplitude (top) and phase (bottom) of the 
  sidebands at the output of a simple Sagnac with no arm cavities.  
  Right: Plots of the amplitude (top) and phase (bottom) of the sidebands at
  the output of a Sagnac with Fabry-Perot arm cavities.  The arms in both 
  cases have length $L=$10\,km.}
\label{fig:mod_sag}
\end{figure}

We now look at the gravitational wave effect on the output of a Sagnac interferometer.  The sideband fields at the output of the detector are given by:
\begin{equation}
b_{\mathrm{out}} = i t_{bs} b_a + r_{bs} b_c
\end{equation}
$b_c$ and $b_a$ refer to the sidebands generated travelling clockwise and anti-clockwise through the interferometer.
\noindent
Travelling clockwise through the interferometer we have:
 \begin{equation}
 b_c = b_c^x + b_c^y R_{cav}(k_0\pm k_g)
 \end{equation}
 where $b_c^x$ and $b_c^y$ refer to the sidebands created in the $x$ and $y$ arms.  $R_{cav}$ is the 
 complex number describing the reflected field from a cavity:
 \begin{equation}
 R_{cav}(k) = r_1 - \frac{T_1 r_2 \exp{(-2ikL)}}{1-r_1 r_2 \exp{(-2ikL)}}
 \end{equation}
 If there is no arm cavity $T_1=1$ and $r_1=0$ and an additional $180^{\circ}$ 
 needs to be added to $R_{cav}$ (mitigating the $90^{\circ}$ phase incurred for 
 each transmission through the input mirror).  
 The sidebands created travelling clockwise through the $y$ arm are given by:
 \begin{equation}
 b_c^y = r_{bs} a_0 (-\alpha_{sb}^{\mathrm{FP}})
 \end{equation}
 The minus refers to the relative phase of the modulation by the gravitational wave.  The sidebands created travelling clockwise through the $x$ arm are given by:
 \begin{equation}
 b_c^x  = r_{bs}  a_0  R_{cav}(k_0) \alpha_{sb}^{\mathrm{FP}} 
 \end{equation} 
So we have:
\begin{equation}
b_c = r_{bs} \ a_0 \ \alpha_{sb}^{\mathrm{FP}} \left[R_{cav}(k_0) - R_{cav}(k_0 \pm k_g) \right]
\end{equation}
\noindent
The sidebands created travelling anti-clockwise through the interferometer are given by:
\begin{equation}
b_a = b_a^x R_{cav}(k_0\pm k_g) + b_a^y
\end{equation}
We have the sidebands created travelling anti-clockwsie through the $x$-arm:
\begin{equation}
b_a^x = it_{bs}  \ a_0 \ \alpha_{sb}^{\mathrm{arm}} 
\end{equation}
The sidebands created travelling anti-clockwise through the 
$y$-arm ($-\alpha_{sb}^{\mathrm{arm}}$ to take into account $h_+$ is out of 
phase by $\pi$ with respect to the $x$ arm):
\begin{equation}
b_a^y = it_{bs}   a_0 R_{cav}(k_0)  (-\alpha_{sb}^{\mathrm{arm}}) 
\end{equation}
\noindent
Which gives the total anti-clockwise sideband field as:
\begin{equation}
b_a = i \ t_{bs} \ a_0 \ \alpha_{sb}^{\mathrm{FP}} \left[ R_{cav}(k_0\pm k_g) - R_{cav}(k_0) \right]
\end{equation}
\noindent
Finally the sidebands at the output of the interferometer are given by:

\begin{equation}
\begin{split}
b_{out} {}& = a_0 \ \alpha_{sb}^{\mathrm{FP}} \left[R_{cav}(k_0) - R_{cav}(k_0\pm k_g)\right] 
\left[ R_{bs} - i^2 T_{bs} \right]  \\
{}& = a_0 \ \alpha_{sb}^{\mathrm{FP}} \left[R_{cav}(k_0) - R_{cav}(k_0\pm k_g) \right] \left[ R_{bs}+T_{bs} \right]
\end{split}
\end{equation}
In figure~\ref{fig:mod_sag} this analytical solution is plotted, as well as the result for a \textsc{Finesse} simulation, for a simple Sagnac and a Sagnac with arm cavities.  The \textsc{Finesse} simulation is detailed in the following kat files:

\clearpage
\begin{multicols}{2}{
\noindent
For a simple Sagnac without arm cavities:
\begin{verbatim}
 l1 1 0 nin

s s0 1 nin n1

const T_ETM 100e-6 

bs BS 0.5 0.5 0 45 n1 ny1 nx1 nout 

s syarm1 10k ny1 ny2 			
bs1 ETMy $T_ETM 0 0 0 ny2 ny3 nytrans dump1
s syarm2 10k ny3 ny4

bs TM 1 0 0 45 ny4 nx4 dump2 dump3

s sxarm1 10k nx1 nx2 			
bs1 ETMx $T_ETM 0 0 0 nx2 nx3 nxtrans dump4
s sxarm2 10k nx3 nx4

fsig sig1 syarm1 1 180
fsig sig1 syarm2 1 180
fsig sig1 sxarm1 1 0
fsig sig1 sxarm2 1 0

ad upper 0 nout

xaxis sig1 f lin 100 50k 400
put upper f $x1
yaxis lin abs:deg




\end{verbatim}

\columnbreak

\noindent
For a Sagnac with arm cavities:
\begin{verbatim}
l l1 1 0 nin

s s0 1 nin n1

const T_ITM 700e-3 
const T_ETM 100e-6 

bs BS 0.5 0.5 0 45 n1 ny1 nx1 nout 

s sy 1 ny1 ny2

bs1 ITMy $T_ITM 0 0 0 ny2 ny3 ny4 ny5
s syarm1 10k ny4 ny6 			
bs1 ETMy $T_ETM 0 0 0 ny6 ny7 ny8 dump1
s syarm2 10k ny7 ny5

bs TM 1 0 0 45 ny3 nx3 dump2 dump3

s sx 1 nx1 nx2

bs1 ITMx $T_ITM 0 0 0 nx2 nx3 nx4 nx5
s sxarm1 10k nx4 nx6 			
bs1 ETMx $T_ETM 0 0 0 nx6 nx7 nx8 dump4
s sxarm2 10k nx7 nx5

fsig sig1 syarm1 1 180
fsig sig1 syarm2 1 180
fsig sig1 sxarm1 1 0
fsig sig1 sxarm2 1 0

ad upper 0 nout

xaxis sig1 f lin 100 50k 400
put upper f $x1
yaxis lin abs:deg
\end{verbatim}
}
\end{multicols}

\begin{thebibliography}{10}

\bibitem{freise2004frequency} A. Freise, G. Heinzel, H. L\"uck,
  R. Schilling, B. Willke, and K. Danzmann, ``Frequency-domain
  interferometer simulation with higher-order spatial modes,''
  Class. Quantum Grav. {\bf 21}, S1067 (2004), the program is
  available at \url{http://www.gwoptics.org/finesse}


\bibitem{Mizuno} J. Mizuno, \emph{Comparison of optical configurations for laser-interferometric gravitational-wave detectors}, PhD. Thesis, University of Hannover (1995).

\end{thebibliography}
\end{document}